\documentclass[aps,prd,reprint,nofootinbib,longbibliography]{revtex4-1}
\usepackage{blindtext}
\usepackage{mathtools}
\usepackage{xcolor}

\usepackage[utf8]{inputenc}
\usepackage[english]{babel}
\usepackage{hyperref}
\hypersetup{
     colorlinks=true,
    linkcolor=blue,
    filecolor=magenta,      
    urlcolor=cyan,
}

\begin{document}
\title{Shining Primordial Black Holes}
\author{Mark P. Hertzberg$^{1*}$, Sami Nurmi$^{2,3**}$, Enrico D. Schiappacasse$^{2,3\dagger}$, Tsutomu T. Yanagida$^{4\ddagger}$}

\affiliation{$^1$Institute of Cosmology, Department of Physics and Astronomy, Tufts University, Medford, MA 02155, USA\\
$^2$ Department of Physics, P.O.Box 35 (YFL), FIN-40014 University of Jyv$\ddot{a}$skyl$\ddot{a}$, Finland\\
$^3$Helsinki Institute of Physics, P.O. Box 64, FIN-00014 University of Helsinki, Finland\\ 
$^4$Tsung-Dao Lee Institute $\&$ School of Physics and Astronomy, Shanghai Jiao Tong University, 
 200240 Shanghai, China }

\begin{abstract}
We study the well-motivated mixed dark matter (DM) scenario composed of a dominant thermal WIMP, highlighting the case of $SU(2)_L$ triplet fermion ``winos", with a small fraction of primordial black holes (PBHs).  After the wino kinetic decoupling, the DM particles are captured by PBHs leading to the presence of PBHs with dark minihalos in the Milky Way today. The strongest constraints for the wino DM come from the production of narrow line gamma rays from wino annihilation in the Galactic Center. We analyse in detail the viability of the mixed wino DM scenario, and determine the constraints on the fraction of DM in PBHs assuming a cored halo profile in the Milky Way. We show that already with the sensitivity of current indirect searches, there is a significant probability for detecting a gamma ray signal characteristic for the wino annihilation in a single nearby dressed PBH when $M_{\text{PBH}} \sim M_{\odot}$, which we refer to as a ``shining black hole".
Similar results should apply also in more general setups with ultracompact minihalos or other DM models, since the accretion of DM around large overdensities and DM annihilation are both quite generic processes. 
\end{abstract}

\maketitle
\section{Introduction}
In the present study, we explore the mixed dark matter (DM) scenario composed of (a dominant) thermal WIMP and primordial black holes (PBHs). We shall especially highlight the case where the WIMP is a kind of fermionic W-boson, namely a ``wino"; although we shall use the name more generally.
The key idea is that PBHs will
unavoidably acquire wino dark matter minihalos after the wino kinetic decoupling. This  would impact indirect searches of DM on the Earth via the enhancement of wino annihilation in such compact astrophysical objects. We would therefore like to explore the possibility that an individual PBH may shine due to emission into photons.

Nowadays, one of the most striking mysteries in modern cosmology and particle physics is
the nature of the bulk of the mass in the universe. Although a wide range of observations, including but not limited to large scale structure, CMB and galactic rotation curves, are very well explained by the inclusion of cold dark matter, its particle physics origin still remains unknown~\cite{Bertone:2016nfn, Freese:2017idy}.

Weakly interacting massive particles (WIMPs) are among the most popular DM candidates in various extensions of the Standard Model (SM) of particle physics. One of the simplest models of weakly interacting DM is the wino~\cite{Cirelli:2005uq} defined by extending the SM via a single electroweak triplet fermion having zero hypercharge~\footnote{A particle physics model which explains DM by including an electroweak triplet fermion was first considered in Ref.~\cite{CHARDONNET199335}.}. This wino multiplet consists in a neutral Majorana fermion, the wino $\chi^0$, and charged fermions $\chi^+\chi-$. The pure wino model
shows the attractive feature of being highly predictive since its mass is the unique relevant parameter to phenomenology (its interactions are determined by the gauge structure of the SM).

On the other hand, within the SM we can have astrophysical objects such as primordial black holes (PBHs) formed in the early Universe~\cite{Hawking:1971ei, Carr:2016drx, Khlopov_2010, Carr:2020erq, PhysRevD.57.6050, Inomata:2017okj, Pi:2017gih, Montero-Camacho:2019jte}. They have been proposed as seeds for cosmic structures~\cite{Carr:2018rid}, the source of gravitational wave events detected by LIGO-Virgo collaboration~\cite{Abbott:2016blz, Sasaki:2016jop} and a possible explanation of the recent NANOGrav results~\cite{Arzoumanian:2020vkk, Kohri:2020qqd, Vaskonen:2020lbd, DeLuca:2020agl, Inomata:2020xad}. 

As a result, the mixed DM scenario mainly composed of thermal wino and 
a small fraction of PBHs appears plausible and well motivated~\footnote{Primordial black holes have also been linked with other DM candidates such as the axion~\cite{PhysRevLett.122.101301, Kitajima:2020kig, Hertzberg:2020hsz}.}. 

In the absence of PBHs, indirect detection related to gamma ray continuum as well as line emission have set tight constraints on wino DM.  Monochromatic $\gamma$-ray lines  come via direct annihilation processes $\chi^0\chi^0 \rightarrow Z\gamma, \gamma\gamma$. Continuous $\gamma$-rays come as secondary decay products of hadrons. Weak gauge bosons $W^+W^-$ coming from the wino annihilation decay into quarks, charged leptons, and neutrinos. After that, the produced quarks undergo fragmentation into various hadrons leading to the formation of stable particles such as protons, electrons and their antiparticles, neutrinos and photons~\cite{Bhattacherjee:2014dya}. Based on the Fermi Gamma-Ray Space Telescope data, constraints on DM from searches of the photon continuum in the Milky Way\textsc{\char13}s satellite dwarf galaxies and the Galactic Center have been reported in Refs.~\cite{Ackermann:2011wa} and ~\cite{Hooper:2012sr}, respectively. Constraints on DM from $\gamma$-ray lines in the Galactic Center have been reported by the Fermi~\cite{Ackermann:2013uma} and HESS collaborations~\cite{Abramowski:2013ax, Abdallah:2018qtu}. 

The strongest limits on wino DM come from photon line searches in the Galactic center. 
The inclusion of a small fraction of PBHs will definitively change the DM distribution and, as a result, indirect detection constraints should be revised. Since PBHs are local overdensities in the DM distribution, they act as seeds for the formation of dark matter structures. Dark matter particles after their kinetic decoupling may be gravitationally bound to the PBHs leading to the formation of minihalos with density spikes~\cite{Eroshenko:2016yve}. Due to the fact that DM annihilation is enhanced by the high particle density, PBHs with minihalos (or dressed PBHs) constitute astrophysical compact objects where wino annihilation is enhanced. Thus, in this scenario we expect in the Milky Way halo today both PBHs with wino minihalos and a smooth wino DM background.

Some previous works have placed bounds on generic WIMP annihilation in the halos of PBHs in terms of the DM particle and PBH masses~\cite{Lacki:2010zf, Aharonian:2008wt, Scott:2009tu, Saito:2010ts,Eroshenko:2016yve, Boucenna:2017ghj, Adamek:2019gns, Bertone:2019vsk}. Other works analyze in particular the thermal wino DM scenario (which comprises the whole DM for a mass $\sim3 \,\text{TeV}$) in the absence of PBHs. Stringent limits based on wino annihilation in dwarf spheroidal galaxies and in the Galactic Center are reported in Refs.~\cite{Bhattacherjee:2014dya, Fan:2013faa} and~\cite{Baumgart:2014saa, Rinchiuso:2018ajn}, respectively. 
Such limits are highly sensitive to uncertainties in the DM density profile, including the details of the core. Furthermore, we will focus on rather heavy wino masses. Altogether, we extend these studies by considering the well-motivated co-dark matter scenario in which heavy dark wino minihalos around PBHs accompany a smooth background of heavy wino DM  with a 
cored DM halo profile, which may avoid the existing bounds~\cite{Cohen:2013ama, Hayashi:2016kcy}.  

In this study, we mainly focus on photon lines emission and analyze the viability of such mixed scenario as well as its possible signals testable by indirect detection surveys on the Earth.

\section{Wino dark matter around PBHs}

The formation of minihalos around PBHs composed of
$\sim 70\, \text{GeV}$-neutralinos was studied first in Ref.~\cite{Eroshenko:2016yve} and then extended for generic WIMPs in
Refs.~\cite{Boucenna:2017ghj, Adamek:2019gns}. In all of these papers, PBHs are tightly constrained to be a small fraction of DM in most parts of the parameter space of interest. In Ref.~\cite{Cai:2020fnq}, dressed PBHs are considered as a source for the Galactic 511 keV line~\cite{1972ApJ...172L...1J, 1975ApJ...201..593H, Kierans:2019aqz}.

The thermal relic wino dark matter (DM) with a heavy mass of $m_{\chi^0}=2.7-3.0$ TeV can explain the current dark matter abundance~\cite{Hisano:2006nn, Hryczuk:2010zi, Beneke:2014hja}. We define $f_{\chi^0} \equiv \Omega_{\chi^0}/\Omega_{\text{DM}}$ and  $f_{\text{PBH}} \equiv \Omega_{\text{PBH}}/\Omega_{\text{DM}}$ as the fraction of dark matter consisting of wino and PBHs, respectively.
Since we are interested in the scenario where PBHs constitute a small fraction of dark matter, e.g. $f_{\text{PBH}} \leq 0.01$, we use the approximation
\begin{equation}
1 = f_{\text{PBH}} + f_{\chi^0} \simeq f_{\chi_0}\,.
\end{equation}
We will fix the wino mass as $m_{\chi^0}=2.8$ TeV for further simplification. Since annihilation mostly takes place in inner shells of minihalos, an accurate calculation of the profile of the wino dark matter around PBHs is important. The wino kinetic decoupling, $t_{\text{KD}}$, and  the PBH mass, $M_{\text{PBH}}$, play a crucial role to determine the density profile of dressed PBHs.   
Let us consider  the PBH formation at the radiation-dominated cosmological stage through  initial density perturbations~\cite{10.1093/mnras/168.2.399, 1975ApJ...201....1C}. The PBH mass is associated with the horizon mass $M_H$ at the time $t_H$ at which the perturbed region crosses the horizon. We may express the PBH mass in terms of the temperature at that time as
\begin{equation}
M_{\text{PBH}} \sim 290 \left( \frac{g_{*}}{10} \right)^{-1/2}\left(\frac{10\,\text{MeV}}{T}\right)^2\,M_{\odot}\,,    
\end{equation}
where we have used $M_{\text{PBH}}=(1/\sqrt{3})^3M_{\text{H}}$ as a simple estimate \cite{1975ApJ...201....1C} and $g_{*}$ is the effective number of relativistic degrees of freedom. 

On the other hand, the kinetic decoupling of the wino takes place at $T_{\text{KD}} \simeq 9.2\, \text{MeV}$ when the annihilation rate is about four times larger than the Hubble parameter~\cite{Kamada:2019eht}. 
Thus, we see that PBHs are formed before the wino kinetic decoupling unless their masses are hundred times larger than the solar mass. As shown in Ref.~\cite{Eroshenko:2016yve}, before the DM kinetic decoupling, the density growth around the central PBH  can safely be neglected. However, once the DM becomes free from the primordial plasma, it may be gravitationally bound to the PBHs and form minihalos with density spikes. The annihilation mainly comes from innermost shells of
minihalos, which are formed between the wino kinetic decoupling in  the early Universe and $z_{eq}$, the redshift for radiation-matter equality. After $z_{eq}$, one has
accretion during the matter-dominated universe via secondary infall accretion~\cite{1985ApJS...58...39B}, but these new layers do not affect the innermost shells already formed very much.

We note that our results are not immediately applicable to astrophysical black holes, but our focus is on PBH instead. Since astrophysical black holes coming from supernovae explosion form later in the Universe (after the first generation of stars), they live in an environment where tidal forces are not negligible, so that even though they
may acquire a minihalo around them, we cannot apply the usual theory of spherical gravitational collapse which assumes isolated black holes~\cite{Mack:2006gz}.

%In contrast, we note that  astrophysical black holes coming from supernovae explosion form later in the Universe (after the first generation of stars).  Even though accretion around
%overdensities is a quite generic process, the annihilation mainly comes from innermost shells of
%minihalos. In our calculation, this part of the minihalo is formed between the wino kinetic decoupling 
%in  the early Universe and $z_{eq}$, the redshift for radiation-matter equality. After $z_{eq}$, one has
%accretion during the matter-dominated universe, but these new layers do not affect the innermost shells already formed very much.  Since astrophysical black holes will form after supernovae collapse, they live in an environment where tidal forces are not negligible, so that even though they
%may acquire a minihalo around them, we cannot apply the usual theory of spherical gravitational
%collapse. As a result, our results are not immediately applicable to astrophysical black holes, but our focus is on PBH instead. 

Suppose that, after the kinetic decoupling, a wino dark matter particle with a velocity $v_i$ is located at a radial distance $r_i$ from a central PBH. At later times, the wino density which ends up in bound orbits around the PBH reads as~\cite{Eroshenko:2016yve}
\begin{equation}
\rho^{\text{PBH}}_{\text{halo}} = \frac{1}{r^2} \int dr_i r_i^2 \rho_i(r_i) \int d^3v_i f(v_i) \frac{2 dt/dr}{T_{\text{orb}}}\,,\label{eq:rhobound}    
\end{equation}
where $\rho_i(r_i)$ is the initial wino DM density, $dt / dr$ is determined by the wino orbit equation in the PBH gravitational potential, $T_{\text{orb}}$ is the classical orbital period and $f(v_i)\propto \text{exp}(-m_{\chi^0}v_i^2/2T)$ is the Maxwell-Boltzmann distribution for the wino velocity.

The wino dark matter density at the kinetic decoupling time $t_{\text{KD}}$ is proportional to the density at the matter-radiation equality $\rho_{\text{eq}}$ as
\begin{equation}
\rho_{\text{KD}}=\left(\frac{\rho_{\text{eq}}}{2}\right)\left(  \frac{a(t_{\text{eq}})}{a(t_{\text{KD}})}\right)^3\simeq 385 \,\text{gr}\, \text{cm}^{-3}\,,    
\end{equation}
where we have used
\begin{equation}
t_{\text{KD}}=\left( \frac{45 m_{\text{pl}}^2}{16 \pi^3 g_{*}(T_{\text{KD}})T^4_{\text{KD}}} \right)^{1/2}\simeq 9\times10^{-3}\,\text{s}\,.    
\end{equation}
The wino temperature and the initial wino DM density in Eq.~(\ref{eq:rhobound}) depend on the relation between the initial radius $r_i$  and the PBH influence radius $r_{\text{inf}}$ (the radius at which the enclosed radiation mass equals the PBH mass) as follows
\begin{equation}
r_{\text{inf}}(t_{\text{KD}})\equiv (8M_{\text{PBH}}t_{\text{KD}}^2/m^2_{\text{pl}})^{1/3}\,.  
\end{equation}
If $r_i \leq r_{\text{inf}}(t_{\text{KD}})$, we have $T=T_{\text{KD}}$ and $\rho_{i}(r_i)=\rho_{\text{KD}}$. In contrast, if the initial radius is larger than the PBH influence radius, we need to scale the wino temperature and initial density as $T=T_{\text{KD}}t_{\text{KD}}/t$ and $\rho_i(r_i)=\rho_{\text{KD}}(t_{\text{KD}}/t)^{3/2}$, where
$r_{\text{inf}}(t)=r_i$ \cite{Eroshenko:2016yve}. In simple words, the influence radius gradually expands until a time $t>t_{\text{KD}}$ such that the wino initial radius is reached by the gravitational action of the PBH.

After numerical evaluation of Eq.~(\ref{eq:rhobound}),
we should compare the density of bound winos with the maximum dark matter density that can be held by an astrophysical object due to annihilation processes, $\rho_{\text{max}}$. Specifically, we have~\cite{PhysRevD.72.103517} 
\begin{widetext}
\begin{equation}
   \rho_{\text{max}} \simeq \frac{m_{\chi^0}}{\langle\sigma v \rangle t_0} \simeq 1.4\times10^{-14}\,\text{gr}\,\text{cm}^{-3}\left( \frac{m_{\chi^0}}{2.8\,\text{TeV}} \right)\left( \frac{8\times10^{-25}\,\text{cm}^{3}\,\text{s}^{-1}}{\langle\sigma v \rangle} \right)\left( \frac{1.4\times10^{10}\,\text{yr}}{t_0} \right)\,,\label{eq:rhomax} 
\end{equation}
\end{widetext}
where $\langle\sigma v \rangle$ is the velocity-weighted cross section of DM annihilation and $t_0$ is the current cosmological time. As we mentioned in the introduction, the total annihilation cross section for the thermal DM wino is 
mostly dominated by the tree level process $\chi^0\chi^0 \rightarrow W^+W^-$. Higher order processes such that  $\chi^0\chi^0 \rightarrow ZZ,Z\gamma,\gamma\gamma$ are suppressed for thermal DM wino. We take $\langle\sigma v \rangle \simeq 8\times10^{-25}\,\text{cm}^3\, \text{s}^{-1}$ for  $m_{\chi^0}=2.8\,\text{TeV}$~\cite{Hryczuk:2011vi}. 

Figure~\ref{Plot1} (top) shows different radial profiles for the wino density in dressed PBHs as a function of PBH masses. The larger the PBH mass, the smaller is the influence radius in units of the Schwarschild radius. Thus, the density profile for the lighter PBHs form a kind of \textit{envelope} to that for the heavier PBHs.
 Since the DM density drops very quickly in minihalos as we departure from the most inner shells, we may estimate the density profile as \cite{Boucenna:2017ghj}
 \begin{equation}
 \rho^{\text{PBH}}_{\text{halo}}(r)\simeq \rho_{\text{max}}\Theta(r^{*}-r)\,, \label{eq:rhoapp} 
 \end{equation}
 where $\rho_{\text{halo}}(r^*)=\rho_{\text{max}}$ as shown in Fig.~\ref{Plot1} (middle). We will apply this approximation later when we calculate the annihilation rate within dressed PBHs. The value of $\tilde{r}^* = r^*/r_g$ for lighter PBHs $M_{\rm PBH}\lesssim 10^{-10} M_{\odot}$ is practically constant, but it quickly decreases as the density profile of heavier PBHs departures from the \textit{envelope} mentioned above. Here $r_g = 2 G_N M_{\text{PBH}}$ is the corresponding Schwarzschild radius.
 
 Here we mention that, after the time of matter-radiation equality, a DM minihalo begins to grow around PBHs  at  radial distances
 $r>r_{\text{inf}}(t_{\text{eq}})$ via secondary infall accretion~\cite{1985ApJS...58...39B}. The density profile associated with this mechanism scales with the radius as $\sim r^{-9/4}$ and the upper density at  $r=r_{\text{inf}}(t_{\text{eq}})$ is many order less than $\rho_{\text{max}}$. Thus, Eq.~(\ref{eq:rhoapp})
remains well justified in describing the innermost shells of the minihalos which are relevant for the DM annihilation processes.
 
\section{Dressed PBHs and $\gamma$-ray lines emission}
As we mentioned in the introduction, for the thermal relic wino the
strongest limits come from searches of photon lines via wino annihilation in the Galactic Center. Consider the photon line emission associated with the processes $\chi^0\chi^0 \rightarrow \gamma \gamma \,\text{and}\, \gamma Z$ and their respective thermally averaged annihilation cross sections $\langle\sigma v \rangle_{\gamma \gamma}$ and  $\langle\sigma v \rangle_{\gamma Z}$. 
\begin{figure}[t!]
\centering
\includegraphics[width=\columnwidth,height=5.9cm]{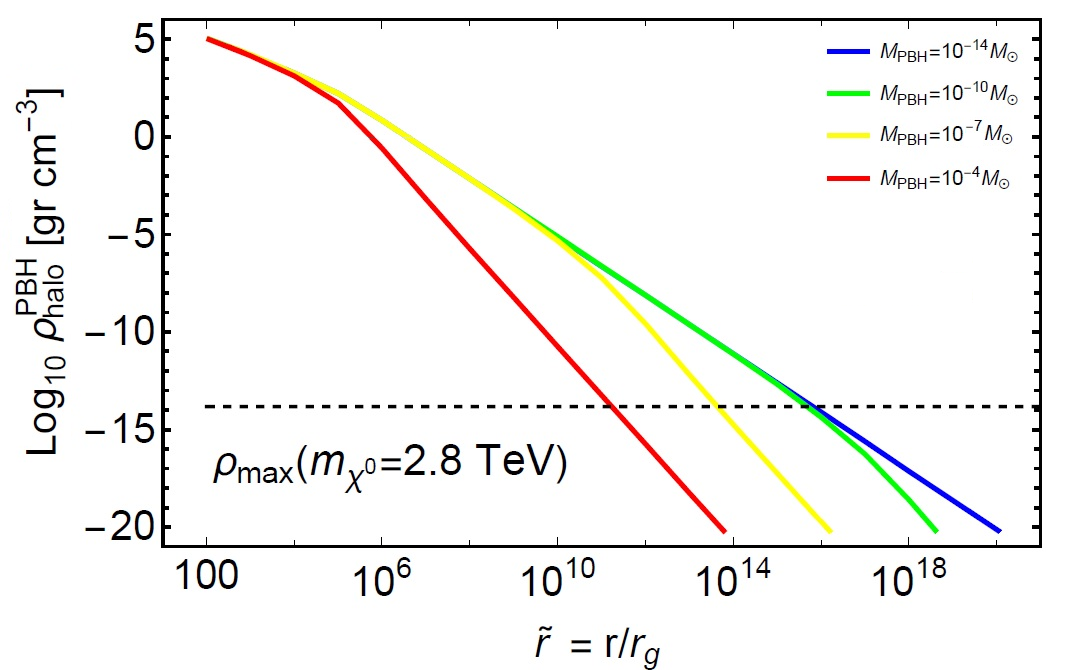}
\includegraphics[width=\columnwidth,height=5.9cm]{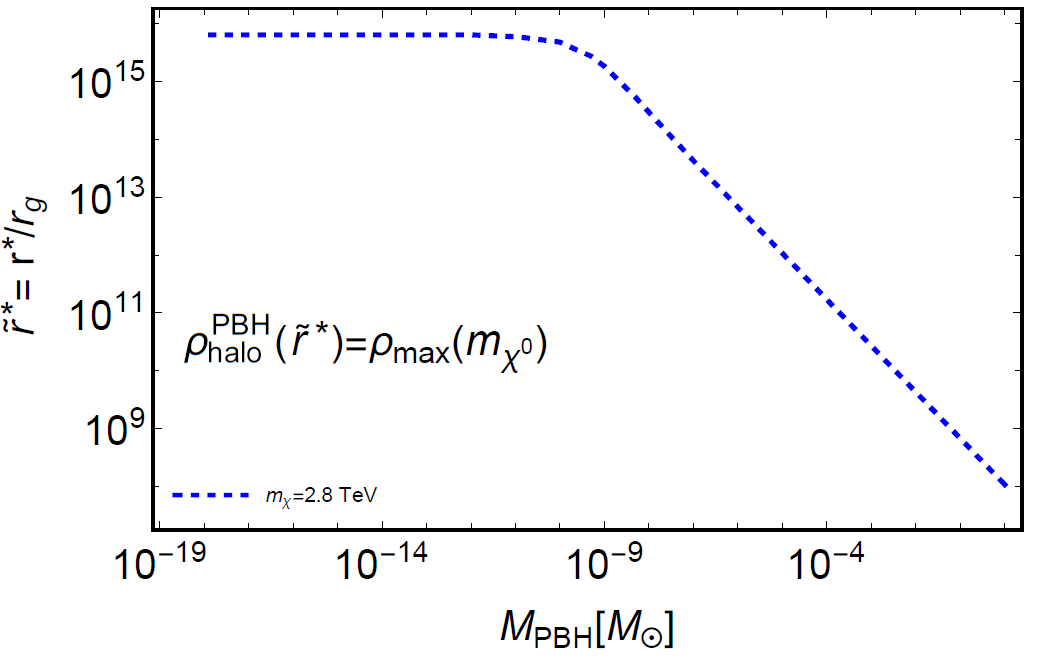}
\includegraphics[width=\columnwidth,height=5.9cm]{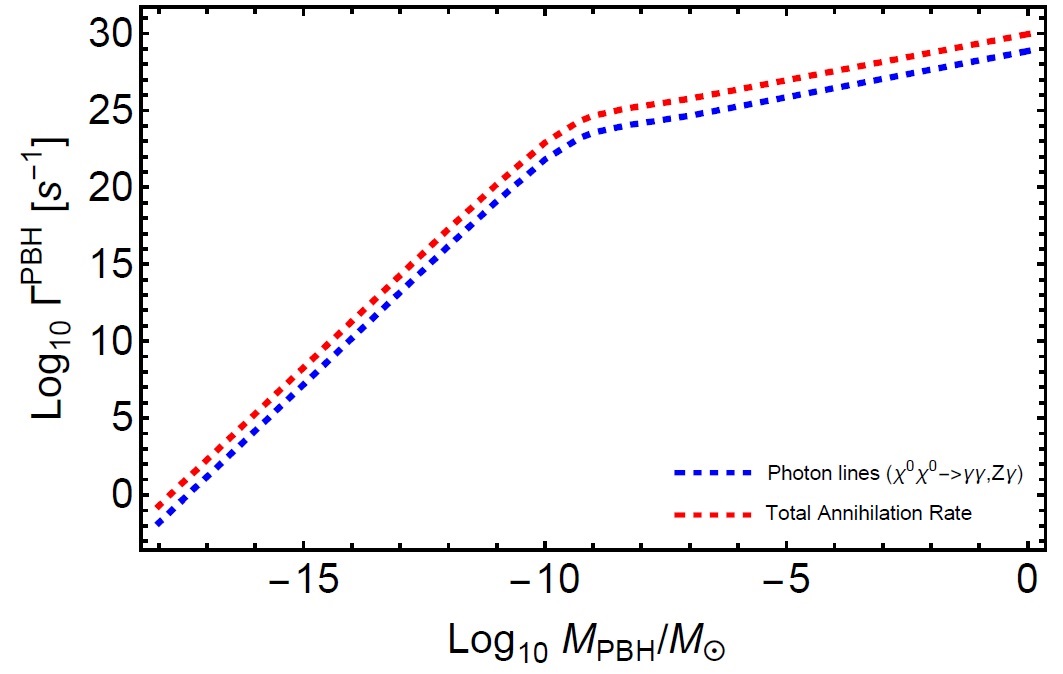}
\caption{
(Top) Wino DM density bound in dressed PBHs in terms of PBH masses. The radius is normalized with the corresponding Schwarzschild radius, $r_g = 2 G_N M_{\text{PBH}}$. The black dashed line indicates the value for $\rho_{\text{max}}$ given in Eq.~(\ref{eq:rhomax}) for the wino case. (Middle) Radii of minihalos in units of $r_g$ around PBHs composed of wino dark matter, $m_{\chi^0}=2.8$ TeV, at which the density hits $\rho_{\text{max}}$ given by Eq.~(\ref{eq:rhomax}). (Bottom) Line annihilation rate in dressed PBHs according to Eq.~(\ref{anntotal}) (blue dashed line). For comparison, the total annihilation rate is shown in the red dashed line.} 
\label{Plot1}
\end{figure}    
The gamma ray spectrum produced by wino annihilation, $dN^{(\gamma)}/dE$, for these two particular channels reads as~\cite{Fan:2013faa}
\begin{align}
&\frac{dN^{(\gamma)}}{dE}(\chi^0\chi^0\rightarrow\gamma\gamma) = 2 \delta\left[E_{\gamma}-m_{\chi^0}\right]\,,\label{gammagamma}\\
&\frac{dN^{(\gamma)}}{dE}(\chi^0\chi^0\rightarrow\gamma Z)=  \delta\left[E_{\gamma}-\left(m_{\chi^0}-\frac{m_z^2}{4m_{\chi^0}}\right)\right]\,.\label{gammaZ}
\end{align}
The difference  between the photon energy in the $\gamma Z$ final state and that in the $\gamma\gamma$ final state is 
\begin{equation}
 \Delta E_{\gamma} \simeq 2\, \text{GeV} \left( \frac{1\,\text{TeV}}{m_{\chi^0}}\right)\,.    
\end{equation}
For H.E.S.S.-like observations typical energy resolutions are at the level $\left(15-20\right)\%$ for individual photons, so that the use of an effective line annihilation cross section into photons $\langle \sigma v \rangle_{\text{line}} = \langle\sigma v \rangle_{\gamma \gamma} + (1/2)\langle\sigma v \rangle_{\gamma Z}$ is well justified.

The line annihilation rate in dressed PBHs depends on the wino density profile and photon line annihilation cross section. Using the approximation  shown in Eq.~(\ref{eq:rhoapp}), we have
\begin{equation}
\small
\Gamma^{\text{PBH}}_{\text{line}}=\langle \sigma v \rangle_{\text{line}} \int d^3r \left(\frac{\rho^{\text{PBH}}_{\text{halo}}(r)}{m_{\chi^0}}\right)^2
\approx \frac{4\pi\langle \sigma v \rangle_{\text{line}} r^{* 3} \rho_{\text{max}}^2}{3 m_{\chi^0}^2}\,, \label{anntotal}  
\end{equation}
where $\rho_{\text{max}}$ is given by Eq.~(\ref{eq:rhomax}) and $\langle \sigma v\rangle_{\text{line}} \simeq 6 \times 10^{-26}\,\text{cm}^3\,\text{s}^{-1}$
is the wino annihilation cross section to line photons for $m_{\chi^0}=2.8 \,\text{TeV}$, including Sommerfeld enhancement (SE) and the resummation of electroweak
Sudakov logarithms at next-to-next-to-leading logarithmic order (NLL)~\cite{Ovanesyan:2014fwa}.   
Figure~\ref{Plot1} (bottom) shows the line annihilation rate for wino into photons (blue dashed line), which is about one order less than the total annihilation rate (red dashed line). The slope break at around $M_{\text{PBH}} \sim 10^{-9}\,M_{\odot}$ is related to the fact that $r^*$ becomes constant in dressed PBHs which hold a central PBH with $M_{\text{PBH}} \lesssim 10^{-10}\,M_{\odot}$. 

We expect significant photon-line signatures coming from annihilation in most inner shells of dressed PBHs. The integrated flux (per solid angle unit) on the Earth coming from a specific sky region in the Milky Way reads as
\begin{equation}
\Phi^{\text{PBH}}_{\text{line}} = \frac{1}{2}\frac{\rho_{\text{E}}r_{\text{E}}}{4\pi} \frac{f_{\text{PBH}}}{M_{\text{PBH}}}\times 2\Gamma^{\text{PBH}}_{\text{line}} \overline{D}(\Delta \Omega)\,,\label{FluxDPBH}   
\end{equation}
where the $1/2$-prefactor is related to the Majorana nature of the wino, the $2$-factor multiplying the photon line annihilation rate refers to the effective number of photons per annihilation and the astrophysical average D-factor is given by~\cite{Cirelli:2010xx}
\begin{align}
\overline{D}(\Delta \Omega)& = \frac{\int_{\Delta \Omega}D d\Omega}{\Delta \Omega}\,,\nonumber\\ 
&=   \frac{1}{\Delta \Omega}\int_{\Delta \Omega} d\Omega \int_{\text{l.o.s}} \frac{\rho_{\text{halo}}^{\text{MW}}(r(s,\theta))}{\rho_{\text{E}}}\frac{ds}{r_{\text{E}}}\,.\label{Dfactor}  
\end{align}
Here $\rho_{\text{halo}}^{\text{MW}}$ is the DM density profile in the Milky Way halo, $\rho_{\text{E}} = 0.3\, \text{GeV}/\text{cm}^3$ is the conventional value for the local DM density and $r_{\text{E}} = 8.5\,\text{kpc}$ is the galactocentric distance of the solar system. The D-factor is calculated along the line of sight (l.o.s.) and the $r$-coordinate, which is
centered in the Galactic Center, reads as 
\begin{equation}
r = \sqrt{r_{\text{E}}^2+s^2-2r_{\text{E}}s\text{cos}(\theta)}\,,    
\end{equation}
with $\theta$ being the angle between the direction of the line of sight and the axis which joins the Earth and the center of the galaxy. For convenience during calculation, this angle may be expressed in terms of the galactic latitude $b$ and longitude $l$ by using $\text{cos}(\theta) = \text{cos}(b)\text{cos}(l)$.

In the absence of PBHs, for a self-conjugate DM annihilation  which generates high-energy photon-lines, the integrated flux (per solid angle unit) on the Earth from a given angular direction $\Delta\Omega$ reads as
~\cite{Salati:2007zz, Cirelli:2010xx} 
\begin{equation}
\Phi_{\text{line}}^{\text{DM}} = \frac{\rho_{\text{E}}^2r_{\text{E}}}{8\pi}\frac{2\langle \sigma v\rangle_{\text{line}}}{m^2_{\chi}} \times \overline{J}(\Delta \Omega)\,,\label{Phi}
\end{equation}
where the astrophysical average J-factor is given by~\cite{Bergstrom:1997fj, Cirelli:2010xx} 
\begin{align}
\overline{J}(\Delta \Omega) &= \frac{\int_{\Delta \Omega}J d\Omega}{\Delta\Omega}\,,\\
&= \frac{1}{\Delta\Omega} \int_{\Delta\Omega}d\Omega\int_{\text{l.o.s.}}\left[\frac{\rho^{\text{MW}}_{\text{halo}}(r(s,\theta))}{\rho_{\text{E}}}\right]^2 \frac{ds}{r_{\text{E}}}\,.
\end{align}
\begin{figure}[t!]
\centering
\includegraphics[width=\columnwidth,height=6.1cm]{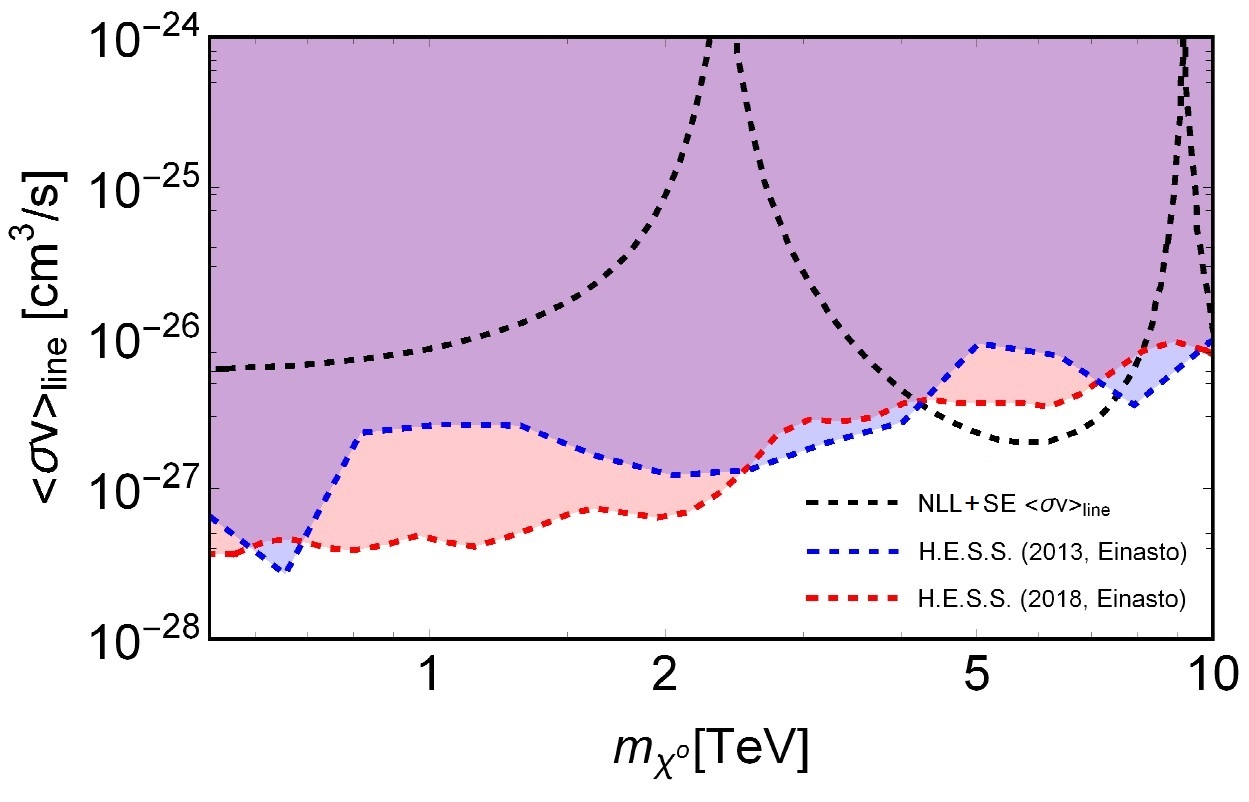}
\includegraphics[width=\columnwidth,height=6.1cm]{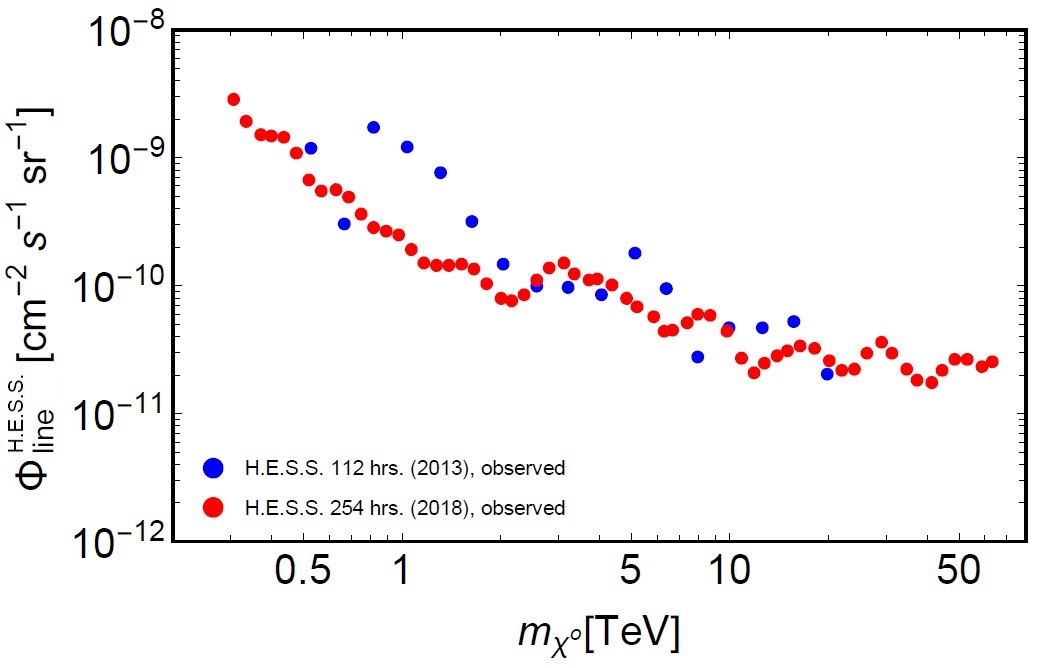}
\caption{ (Top) Upper limits for the velocity-weighted cross section for DM annihilation into two photons. Blue and red dashed lines refer to H.E.S.S. four years data~\cite{Abramowski:2013ax} and ten years data~\cite{Abdallah:2018qtu}, respectively, both for an Einasto profile. The dashed black line corresponds to the NLL+SE velocity-weighted cross section for wino annihilation to line photons from $ \gamma\gamma$ and $\gamma Z$~\cite{Ovanesyan:2014fwa}. (Bottom) Upper limits on $\gamma$-ray flux from monochromatic
line signatures derived from the Galactic Center region for the case of H.E.S.S. four years data~\cite{Abramowski:2013ax} (blue points) and ten years data~\cite{Abdallah:2018qtu} (red points).}  
\label{Plot2}
\end{figure}Note that while this flux depends on the square of the DM halo profile along the line of sight through the J-factor, the flux coming from dressed PBHs depends only on one power of the DM halo profile. Thus, $\Phi^{\text{DM}}_{\text{line}}$ is more sensitive to the distribution of DM in the galaxy than $\Phi^{\text{PBH}}_{\text{line}}$~\footnote{In the standard scenario, the J-factor naturally appears in the calculation of the flux associated with DM annihilation and, for this reason, depends on $\left(\rho^{\text{MW}}_{\text{halo}}\right)^2$. By contrast, the D-factor, which is proportional to one power of $\rho^{\text{MW}}_{\text{halo}}$, appears during the study of the flux associated with decaying DM, where only one parent particle is involved. In our particular case, the DM self-annihilation occurs within dressed PBHs making the flux expression proportional to $\left(\rho^{\text{PBH}}_{\text{halo}}\right)^2$ through the annihilation rate, Eq.~(\ref{anntotal}). However, as the number density of PBHs tracks the DM halo density profile, the flux is proportional to only one power of $\rho^{\text{MW}}_{\text{halo}}$ leading to a D-factor in the final expression, Eq.~(\ref{FluxDPBH}).}. 

Figure~\ref{Plot2} (top) shows the upper limits
for the velocity-weighted cross section of DM annihilation into two photons from H.E.S.S. observations of the Galactic Center region.  Blue and red dashed lines refer to H.E.S.S.  four years~\cite{Abramowski:2013ax} and ten years data~\cite{Abdallah:2018qtu}, respectively, both for an Einasto profile. The dashed black line refers to the (NLL+SE) $\langle \sigma v \rangle_{\text{line}}$ for wino annihilation to line photons~\cite{Ovanesyan:2014fwa}. 

Limits from H.E.S.S. shown in Fig.~\ref{Plot2} (top) indicate that the thermal relic wino is ruled out by more than one order of magnitude assuming the Einasto profile. However, the photon flux measured by the H.E.S.S. experiment is proportional to the J-factor along the line of sight, 
and the current astrophysical uncertainties in
the halo profile are large enough to evade H.E.S.S. limits.

The H.E.S.S. region of interest is defined as a circle of $1^o$ radius centered on the Galactic Center with the exclusion of the Galactic plane (the Galactic latitudes are restricted as $|b|>0.3^o$)~\cite{PhysRevLett.106.161301}. 
The exclusion curves shown in Figure~\ref{Plot2} (top) extracted from Refs.~\cite{Abramowski:2013ax, Abdallah:2018qtu} assume an Einasto DM profile~\cite{1965TrAlm...5...87E}
normalized to the local DM density, $\rho_{\odot} = 0.39\,\text{GeV}\,\text{cm}^{-3}$ at a distance of 8.5 kpc from the Galactic Center~\cite{Catena:2009mf}. Such profile is parametrized as follows~\cite{Pieri:2009je}
\begin{equation}
\rho^{\text{MW}}_{\text{Einasto}}(r)=\rho_s\, \text{exp}\left[-\frac{2}{\alpha_s}\left(\left(\frac{r}{r_s}\right)^{\alpha_s}-1 \right) \right]\,,
\label{eq:NFWE}
\end{equation}
where $r$ is the radial distance from the center of the galaxy, $\rho_s = 0.079 \text{ GeV cm}^{-3}$ and $r_s = 20\text{ kpc}$ are the scale density and radius, respectively, and $\alpha_s = 0.17$ defines how steep the DM profile is. 

The Einasto profile~\cite{1965TrAlm...5...87E} (as well as the Navarro-Frenk-White profile, NFW~\cite{1996ApJ...462..563N}) is a standard example of a cusped density profile which is supported by DM particle simulations without baryonic physics effects. The inclusion of baryonic matter in simulations appears to lead to the formation of a ``cored" profile, where the short-distance cusp is flattened out. While cosmological simulations from the FIRE (Feedback In Realistic Environments) project in Ref.~\cite{Chan:2015tna} reported DM profiles slightly shallower than a NFW profile having cores of  $\sim 1 \text{ kpc}$,  cosmological hydrodynamical simulations of Milky Way-sized halos including tuned star formation rate and supernovae feedback performed in Ref.~\cite{Mollitor:2014ara} produce cores with a typical size of $\sim 5 \text{ kpc}$.  In addition, dynamical models of the Milky Way Galactic bulge, bar and inner disk using the Made-to-Measure method reported in Ref.~\cite{2017MNRAS.465.1621P} favor a shallow cusp or a core in the bulge region.

A cored profile can be parametrized using the 
profile defined in Eq.~(\ref{eq:NFWE}). We define a cutoff-Einasto profile as
\begin{equation}
\rho_{\text{Cutoff-Einasto}}^{\text{MW}}(r) = 
\begin{cases}
\rho^{\text{MW}}_{\text{Einasto}}(r_c) \,\,\text{for}\,\, r \leq r_c\,, \\
\,\hspace{3 cm}\\
\rho^{\text{MW}}_{\text{Einasto}}(r)\,\,\text{for}\,\, r > r_c\,.
\end{cases}
\label{eq:NFWEcutoff}
\end{equation}
 By calculating the variation of the J-factor in the region of interest and varying the core size, we may evaluate the minimum core size needed to make wino DM ($m_{\chi^0}=2.8\,\text{TeV}$) compatible with H.E.S.S. limits. 
 
In the mixed DM scenario that we are studying, we expect to have dressed PBHs as well as a smooth DM background.  In Ref.~\cite{Hertzberg:2019exb} some of us performed a 
detailed analysis about the amount of DM which can be today in the Galactic halo in the form of minihalos around PBHs. Based on previous numerical simulations~\cite{Mack_2007}, 
we expect that dressed PBHs, consisting of the inner core described by Eq. (\ref{eq:rhoapp}) and outer parts from the secondary infall accretion, typically constitute a fraction of DM of the order of 
$10^2 f_{\text{PBH}}$. Then, 
the fraction of DM in the smooth background may be estimated as $\sim (1-10^2f_{\text{PBH}})$. Putting all together,
the integrated fluxes over the region of interest
coming from dressed PBHs and the smooth background ($\Phi_{\text{line}}^{\text{smooth}}$) should
satisfy the following inequality:
\begin{equation}
     f_{\text{PBH}}\overline{\Phi}_{\text{line}}^{\text{PBH}} +  (1-10^2f_{\text{PBH}})^2\overline{\Phi}_{\text{line}}^{\text{smooth}} \leq  \Phi_{\text{line}}^{\text{H.E.S.S.}}\,,\label{fluxineq}
\end{equation}
where
\begin{align}
&\Phi_{\text{line}}^{\text{PBH}}=f_{\text{PBH}}\overline{\Phi}_{\text{line}}^{\text{PBH}}\,,\\
&\Phi_{\text{line}}^{\text{smooth}}= (1-10^2f_{\text{PBH}})^2\overline{\Phi}_{\text{line}}^{\text{smooth}}\,.
\end{align}
Here $\Phi_{\text{line}}^{\text{PBH}}$ is given by Eq.~(\ref{FluxDPBH}),
$\Phi_{\text{line}}^{\text{H.E.S.S.}}$ is the upper limit on $\gamma$-ray flux of monochromatic line signatures from H.E.S.S. observations shown in Fig.~\ref{Plot2} (bottom) and
$\Phi_{\text{line}}^{\text{smooth}}$ is the expected 
flux from monochromatic line signatures associated with the cross section for wino annihilation to line photons in Fig.~\ref{Plot2} (top, dashed black line) using a cored DM halo profile and a partial fraction of DM in wino (smooth background).

To estimate upper limits on the fraction of DM in PBHs, we only consider the  updated results from H.E.S.S. Collaboration (2018)~\cite{Abdallah:2018qtu}. We take the Einasto profile with core sizes ranging as $3\, \text{kpc} \leq r_c \leq 10\, \text{kpc}$. The smallest size core in this range is somewhat larger than the minimum size core that a thermal relic wino needs to avoid H.E.S.S. limits (in the absence of PBHs). At that minimum size core, we have $\Phi_{\text{line}}^{\text{smooth}} = \Phi_{\text{line}}^{\text{H.E.S.S.}}$ for $f_{\text{PBH}}=0$ in Eq.~(\ref{fluxineq}).      

\begin{figure}[t!]
\centering
\includegraphics[width=\columnwidth,height=6.1cm]{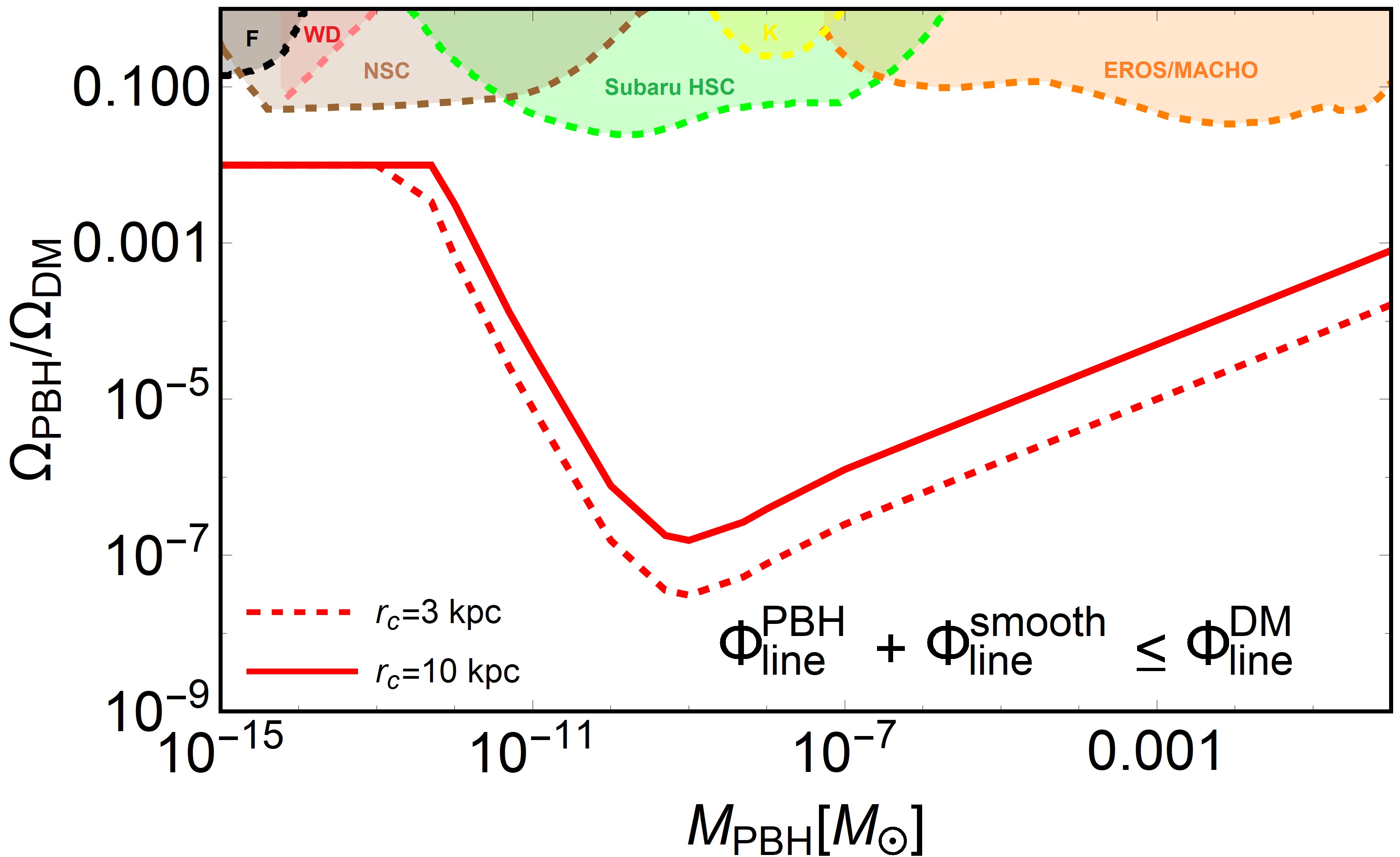}
\caption{ Upper limits for the fraction of DM in PBHs for the mixed DM scenario with thermal relic wino and PBHs. We use upper limits on $\gamma$-ray flux from monochromatic line signatures from the Galactic Center reported by H.E.S.S. Collaboration in Ref.~\cite{Abdallah:2018qtu}. The dark matter halo is parametrized using a cutoff-Einasto profile with a core size $r_c = 3\,\text{kpc}$ (dashed red line) and $10\,\text{ kpc}$ (red solid line). The color shaded regions show PBH constraints from femtolensing (F)~\cite{2012PhRvD..86d3001B}, white dwarfs in our local galaxy (WD)~\cite{Graham:2015apa}, neutron-star capture (NSC)~\cite{Capela:2013yf}, Subaru HSC microlensing (Subaru HSC)~\cite{Niikura:2017zjd},
Kepler microlensing of stars (K)~\cite{PhysRevLett.111.181302} and
EROS/MACHO microlensing (EROS/MACHO)~\cite{Tisserand:2006zx}.}  
\label{Plot3}
\end{figure}

Figure~\ref{Plot3} shows upper limits for the fraction of DM in PBHs when 
the DM halo profile is described by a cutoff-Einasto profile with a size of
$r_c = 3 \text{ kpc}$ (red dashed line) and $10 \text{ kpc}$ (red solid line).
In addition, PBH constraints associated with femtolensing~\cite{2012PhRvD..86d3001B}, microlensing~\cite{Niikura:2017zjd, PhysRevLett.111.181302, Tisserand:2006zx}, white dwarfs galaxy~\cite{Graham:2015apa} and neutron-star capture~\cite{Capela:2013yf} are shown. Here we have imposed the condition $f_{\text{PBH}} \leq 0.01$ to be consistent with our assumptions. 
Primodial black holes with a mass $\sim 10^{-9}M_{\odot}$ undergo the strongest constraint, which is expected. The reason is the following. The scaled radius $\tilde{r}^* = r^*/r_g$ at which $\rho_{\text{halo}}^{\text{PBH}}(r^*) = \rho_{\text{max}}$ reaches its maximum value at $M_{\text{PBH}} \lesssim 10^{-9}M_{\odot}$ becoming independent of the PBH mass. In that regime, the
annihilation rate scales with the PBH mass as $\sim{M_{\text{PBH}}}^3$ so that the flux coming from dressed PBHs scales as $\sim{M_{\text{PBH}}}^2$. 

We note that the upper limit on  $f_{\text{PBH}}$ in Fig.~\ref{Plot3} for dressed PBHs with $M_{\text{PBH}} \sim M_{\odot}$ and the case of a cutoff-Einasto profile with a size core of $r_c=10\, \text{kpc}$, $f_{\text{PBH}} \sim 10^{-3}$, is consistent with the needed fraction in PBHs with $M_{\text{PBH}} = \mathcal{O}(1-10) M_{\odot}$ to explain the recent NANOGrav results as shown one of us in Ref.~\cite{Inomata:2020xad}.

In our estimates, we are only considering the line spectrum 
from wino annihilation to $\gamma\gamma$ and $\gamma Z$ final states.
However, $\gamma$-ray photons coming from other annihilation states
will unavoidably add to these line photons if the energy of these
photons changes by an amount less than the energy resolution of the
telescope. The precision wino photon spectrum
~\cite{Baumgart:2017nsr, Baumgart:2018yed} is used in Ref.~\cite{Rinchiuso:2018ajn}
to include endpoint and continuum contributions to constrain
thermal wino DM using a mock
H.E.S.S.-I-like observation of the Galactic Center.
Even though the line contribution to the wino
annihilation spectrum dominates the overall limits in the
TeV mass range, the endpoint and continuum effects become more 
relevant as the wino mass increases. These effects would introduce
an order of one correction in our analysis. 

\section{Isolated dressed PBHs and possibility of detection}

Here we analyze the flux coming from isolated dressed PBHs located at a certain
distance from the Earth. The $\gamma$-ray flux integrated above a threshold $E_{\text{R}}$ from a dressed PBH, treated as a point source and located at a distance $d$, is given by~\cite{Scott:2009tu}
\begin{align}
&\Phi^{\text{PBH}}_{\text{PS}}(E_{\text{R}},d) =\,\nonumber\\
&\sum_k \int_{\text{E}_{\text{R}}}^{m_{\chi^0}} \frac{dN_k^{(\gamma)}}{dE}dE\frac{\langle\sigma v\rangle_k}{8\pi d^2}\int d^3r \left(  \frac{\rho_{\text{halo}}^{\text{PBH}}(r)}{m_{\chi^0}}\right)^2\,,\label{fluxps}     
\end{align}
where the sum runs over the $k$-annihilation channels. 

For the case of photon lines emission, $dN_k^{(\gamma)}/dE$ is given by Eqs.~(\ref{gammagamma}) and (\ref{gammaZ}) and the corresponding velocity-weighted cross sections may be combined in an effective anihilation cross section into photons, $\langle \sigma v \rangle_{\text{line}}$, as we did before. However,  the situation is more involved for the case of the continuous energy spectrum. Photons come mostly from fragmentation 
of hadronic final states in the processes: $\chi^0\chi^0\rightarrow W^+W^-$ (dominant channel) and $\chi^0\chi^0\rightarrow ZZ$. The corresponding two fragmentation functions, $dN^{(\gamma)}_{W^+W^-}/dE$  and $dN^{(\gamma)}_{ZZ}/dE$,
are obtained in Ref.~\cite{Hisano:2004ds} for a wino mass of $\mathcal{O}(\text{TeV})$ via the HERWIG Monte Carlo Code~\cite{Corcella:2000bw}. Both fragmentation functions quickly drop as the energy increases being well fitted by the function  
\begin{equation}
    \frac{dN_k^{(\gamma)}}{dx} = a\,\frac{\text{exp}(b x)}{(x^{3/2}+c)}\,,
\end{equation}
where $a, b,c$ are constants and $x=E/m_{\chi^0}$. The corresponding  thermally averaged annihilation cross sections, $\langle \sigma v\rangle_{W^+W^-}$ and $\langle \sigma v\rangle_{ZZ}$, are calculated in Ref.~\cite{Hryczuk:2011vi} up to one-loop radiative corrections and including electroweak Sommerfeld effect. 

First, let us consider $d = 100\, \text{pc}$ as a canonical value for the distance of the isolated dressed PBH. Considering that the radial flux profile of dressed PBHs is highly dominated by the innermost part of the minihalo, they may be treated as point sources at that distance to current/future experiments~\footnote{A similar argument holds for ultracompact minihalos grown from initial density perturbations. See, for example, Ref.~\cite{Scott:2009tu}.}.  The angular size of these compact objects  is $\lesssim 10^{-6}$ radians for $M_{\text{PBH}} \leq M_{\odot}$ at distances $d \gtrsim 10\, \text{pc}$, under the approximation performed in Eq.~(\ref{eq:rhoapp}).  For an energy scale $\lesssim 3\, \text{TeV}$, the typical instrumental angular resolution for HESS and the planned Cherenkov Telescope Array (CTA)~\cite{2011ExA....32..193A} are $\sim10^{-3}$ and $\sim10^{-4}$ radians, respectively~\cite{ACHARYA20133}.

\begin{figure}[t!]
\centering
\includegraphics[width=\columnwidth,height=6.1cm]{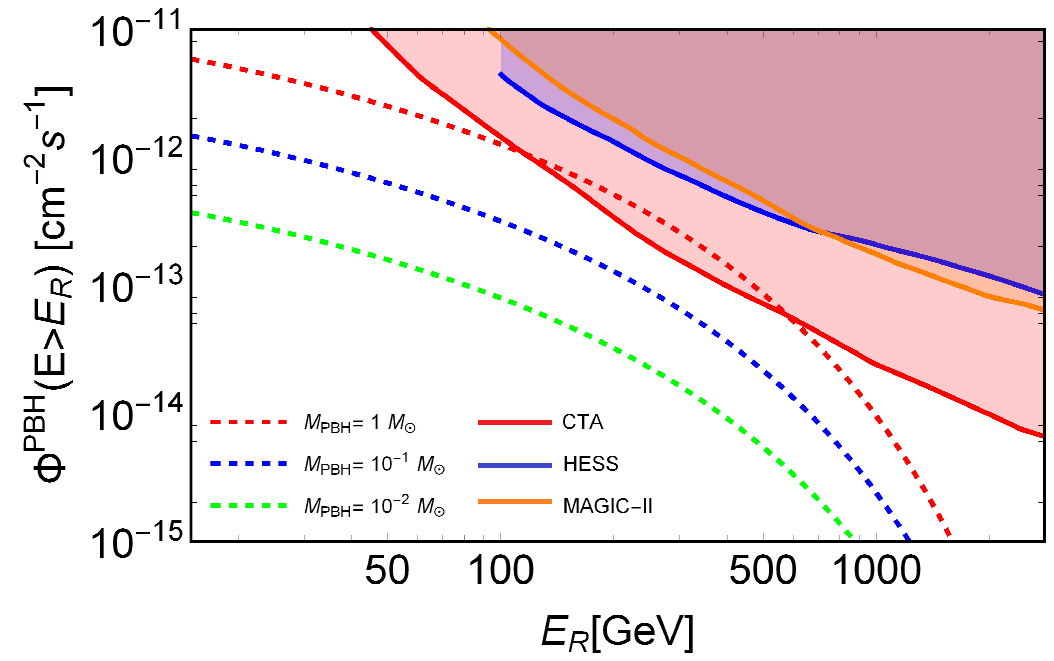}
\caption{Fluxes from an isolated dressed PBH located at a distance $d=100\, \text{pc}$ as function of the energy threshold of the observing experiment. Dashed red, blue and green lines correspond to dressed PBHs with a central PBH with $M_{\text{PBH}}= (1, 10^{-1}, 10^{-2})\,M_{\odot}$, respectively. Red, blue and orange solid lines refer to the integral sensitivity ($E > E_R$) of CTA, HESS and MAGIC-II, respectively, for 50 hours of observations~\cite{ACHARYA20133}.}  
\label{Plot4}
\end{figure}

Figure~\ref{Plot4} shows the expected fluxes from an isolated dressed PBH located at 100 pc as a function of the energy threshold together with the detectable regions associated with the integral flux sensitivity of air Cherenkov telescopes (IACTs~\cite{HILLAS201319}): HESS, MAGIC-II~\cite{Magic} and CTA (see Fig. 8 of Ref.~\cite{ACHARYA20133}). 

In the mass range of our interest for the central PBH, dressed PBHs located at 100 pc are not detectable for current IACTs such as HESS and MAGIC-II. Interesting enough, 
 the flux for a dressed PBH with a central PBH of $M_{\text{PBH}} = M_{\odot}$ is larger than the integral flux sensitivity of the next-generation CTA in one part of the 0.1 TeV scale of the energy threshold.
 
Now, let us consider the characteristic distance $\overline{d}$ at which a particular dressed PBH may be detected as a $\gamma$-ray point source given a certain instrumental sensitivity. This characteristic distance  is calculated by equating the point source flux in Eq.~(\ref{fluxps}), $\Phi_{\text{PS}}^{\text{PBH}}(E_{\text{R}},\overline{d})$, with the corresponding instrumental sensitivity. 
For the continuous energy spectrum,
dressed PBHs with a central PBH  with a mass of $10^{-3} M_{\odot} \lesssim M_{\text{PBH}} \lesssim 1 M_{\odot}$ located at a distance $13\, \text{pc} \lesssim \bar{d} \lesssim 53\, \text{pc}$ should be detectable by HESS with an integral flux sensitivity $\Phi(E> 100\, \text{GeV}) \simeq 4.4\times10^{-12}\,\text{cm}^{-2}\text{s}^{-1}$~\footnote{A more accurate setup may be accomplished  by performing a likelihood analysis taking account the differential energy spectrum
from wino annihilation in each energy bin.}. 
For the photon line case, the large IACTs show high sensitivity in the TeV-scale $\gamma$-rays. For example, MAGIC may reach $10^{-14}\text{cm}^{-2}\text{s}^{-1}$ at the TeV scale in the northern hemisphere and CANGAROO III~\cite{Tsuchiya:2004wv} and HESS may reach $10^{-13}\text{cm}^{-2}s^{-1}$ in the southern hemisphere. The characteristic distance for detection via photon lines range as  $22\, \text{pc} \lesssim \overline{d} \lesssim 86\, \text{pc}$ for $10^{-3} M_{\odot} \lesssim M_{\text{PBH}} \lesssim 1 M_{\odot}$, respectively, when 
an instrumental sensitivity of $10^{-13}\,\text{cm}^{-2}\text{s}^{-1}$ is considered.

We may compare the characteristic distances $\overline{d}$ with 
the expected distance $d$ to the nearest PBH, $d \approx [3M_{\text{PBH}}/(4\pi \rho_{\text{PBH}})]^{1/3}$, where $\rho_{\text{PBH}}$ is the PBH density.
Considering that $\rho_{\text{PBH}} = f_{\text{PBH}} \rho_{\text{DM}}$, we may obtain an upper bound for the fraction of DM in PBHs as
\begin{equation}
    f_{\text{PBH}} \lesssim \frac{3 M_{\text{PBH}}}{4\pi \rho_{\text{DM}}\overline{d}^3}\,,\label{fPBHdbar}
\end{equation}
where $\overline{d}$ is solved by setting Eq.~(\ref{fluxps}) equal to the instrumental sensitivity, and we take $\rho_{\text{DM}} = 0.39\,\text{GeV}\text{cm}^{-3}$ as the DM density in the solar vicinity.

\begin{figure}[t!]
\centering
\includegraphics[width=\columnwidth,height=6.1cm]{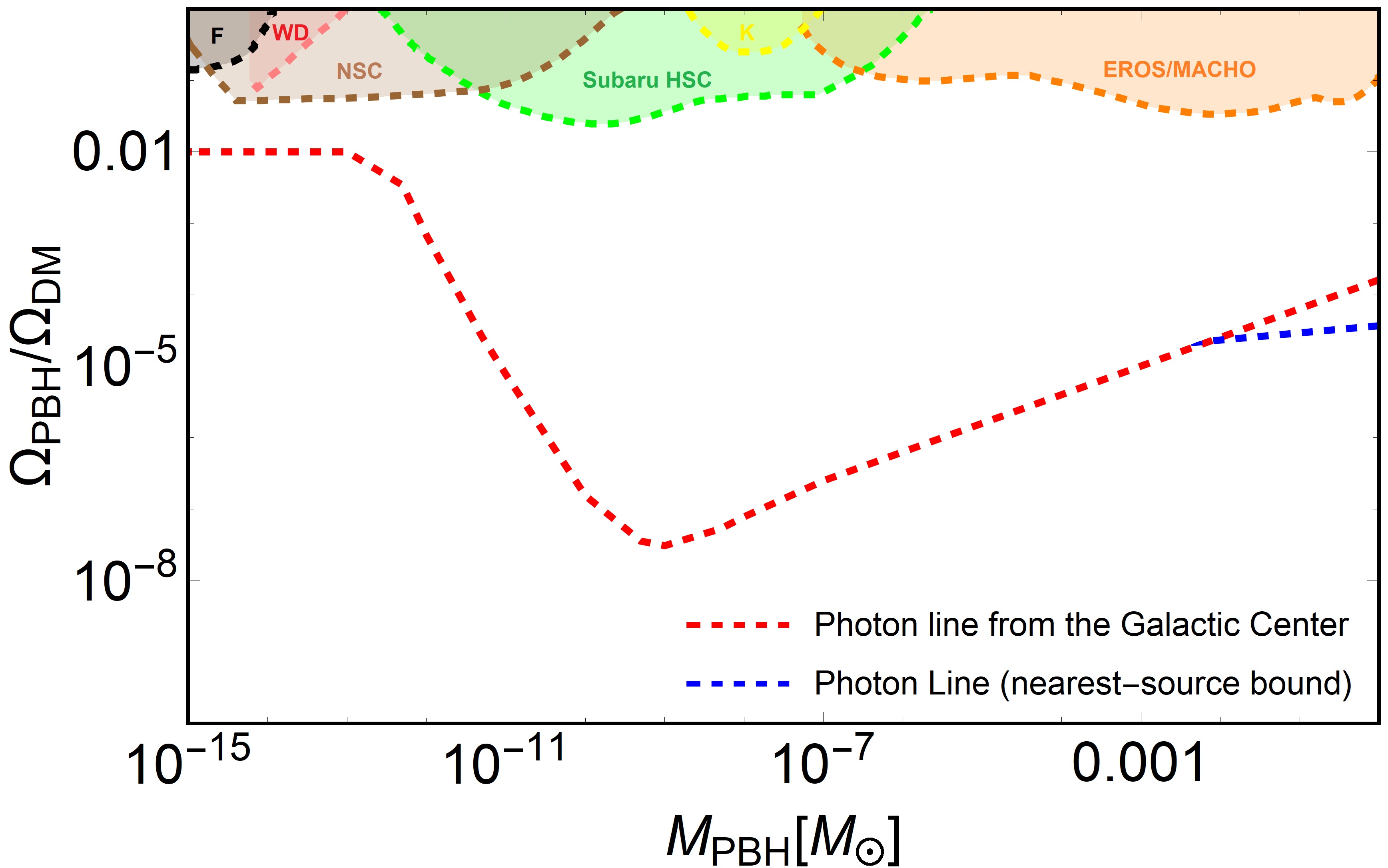}
\caption{ Upper limits for the fraction of DM in PBHs for monochromatic line signatures from the Galactic Center using a cutoff-Einasto profile with $r_c = 3 \text{kpc}$ (red dashed line in Fig. 3) and individual dressed PBHs in the region of interest (blue dashed line). We also show PBH constraints from femtolensing (F), white dwarfs (WD), neutron-star capture (NSC), and microlensing (Subaru HSC, EROS/MACHO and K) in the shaded regions (see caption in Fig. 3).}  
\label{Plot5}
\end{figure}

While for the continuum energy spectrum the nearest-source bound Eq.~(\ref{fPBHdbar}) on $f_{\text{PBH}}$ is weaker than that obtained from photon line emission in the Galactic Center, the nearest-source bound from $\gamma$-ray lines is slightly stronger for $M_{\text{PBH}} \gtrsim 10^{-3}\,M_{\odot}$. Figure~\ref{Plot5} shows the nearest-source bound in the region of interest for the photon line case (blue dashed line) and the Galactic Center bound obtained in the previous section (red dashed line in Fig. 3).~\footnote{A similar situation was reported in Ref.~\cite{Carr:2020erq} for dressed PBHs with massive central PBHs ($M_{\text{PBH}} \gtrsim 10^{6} M_{\odot}$) and minihalos of generic WIMPs. The nearest-source bound associated with dressed PBHs with a central PBH mass larger than a critical value resulted to be stronger than the extragalactic background bound. However, this critical value is so large that the individual bound is placed well outside the incredulity limit.}   

Finally, we estimate the probability of observing one or more dressed PBH at a given distance $d$~\cite{2009PhRvL.103p1301B}.  We assume the number of dressed PBHs in the Milky Way halo equals the corresponding total number of PBHs,  $N^{\text{PBH}}_{\text{halo}} = f_{\text{PBH}} M_{\text{DM}}^{\text{MW}}/M_{\text{PBH}}$, where $M^{\text{MW}}_{\text{DM}}$ is the  
DM mass in Milky Way. The probability that any given one of these dressed PBHs is within a distance $d$ from us, $P(\leq d)$, may be estimated by the ratio between the enclosed DM mass inside a sphere centered on the Sun, $M(\leq d)_{\text{DM}}$, and $M^{\text{MW}}_{\text{DM}}$. Note that as  $\rho_{\rm PBH} = f_{\rm PBH}\rho_{\rm DM}$, the mass ratio just equals to $N^{\text{PBH}}(\leq d)/  N^{\text{PBH}}_{\text{halo}}$, where $N^{\text{PBH}}(\leq d)$ is the expected number of PBHs in the sphere of radius $d$.

In detail, we have
\begin{align}
&P(\leq d) = \frac{M(\leq d)_{\text{DM}}}{M_{\text{DM}}^{\text{MW}}}\,,\\
&= \frac{\int\,d\Omega \int_0^{d}s^2 \rho_{\text{halo}}^{\text{MW}}(r(s,b,l))ds }{
M^{\text{MW}}_{\text{DM}}
%4\pi \int_0^{R_{200}}\rho_{\text{halo}}^{\text{MW}}(r) r^2dr
}
\,.\label{probone}    
\end{align}
We may estimate the probability that at least $k$ dressed PBHs are within
$d$, $P(\leq d)_{\geq k}$, by using the cumulative distribution function for a binomial distribution with parameters $N_{\text{PBH}}$ and $P(\leq d)$. 
%Here we take the total number of dressed PBHs in the Milky Way, $N^{\text{PBH}}_{\text{halo}}$, to be equal to the corresponding number of PBHs, so that   $N^{\text{PBH}}_{\text{halo}} = f_{\text{PBH}} M_{\text{DM}}^{\text{MW}}/M_{\text{PBH}}$.
Since the probability that less than $k$ dressed PBHs are within
$d$ is given by
\begin{align}
 &P(\leq d)_{<k}=\nonumber\\
 &\sum_{i=0}^{k-1}  {N^{\text{PBH}}_{\text{halo}}\choose i}   P(\leq d)^i \left[1-P(\leq d)\right]^{N^{\text{PBH}}_{\text{halo}}-i}\,,
\end{align}
we readily have
\begin{equation}
P(\leq d)_{\geq k} = 1 - P(\leq d)_{< k}\,.
\end{equation}
For the case $k=1$, i.e. the probability that at least one dressed PBH is located within $d$, we have the simple expression 
\begin{equation}
P(\leq d)_{\geq 1} = 1-\left[1-P(\leq d)\right]^{N^{\text{PBH}}_{\text{halo}}}\,.\label{probfull}
\end{equation}
%Then, the total probability of one or more dressed PBHs is located a distance $d$
%reads 
%\begin{equation}
%P^{\text{PBH}}_{\text{obs}} = x P(d<\overline{d})_{\geq 1}.
%\end{equation}
If $M(d<\overline{d}) \ll M_{\text{DM}}^{\text{MW}}$, which is the case of our interest, we may Taylor expand Eq.~(\ref{probfull}) to obtain
\begin{equation}
P_{\text{obs}}^{\text{PBH}} \simeq f_{\text{PBH}} \frac{M(d<\overline{d})}{M_{\text{PBH}}}\,,\label{Pobs}    
\end{equation}
which is independent of the total DM mass in the Milky Way.

Consider dressed PBHs having a central PBH with $M_{\text{PBH}}=M_{\odot}$
with an associated upper bound of $f_{\text{PBH}} \lesssim 3.7 \times 10^{-5}$ 
(dashed blue line in Fig.~\ref{Plot5}). We calculate the
DM mass enclosed in a sphere of radius $\overline{d}$ centered on the Sun
in Eq.~(\ref{probone}) by using the cutoff-Einasto profile with a core size of $r_c=3\,\text{kpc}$ defined in the previous section.
The probability that at least one of these compact objects is located at 100 pc leading to the integral flux shown in Fig.~\ref{Plot4} is about $80\%$~\footnote{Note that if we use the upper limit on $f_{\text{PBH}}$ for $M_{\text{PBH}}\sim M_{\odot}$ derived from wino annihilation in the Galactic center for the case of a cutoff-Einasto profile with $r_c= 10\, \text{kpc}$ (red solid line in Fig.~\ref{Plot3}), the number of dressed PBHs per unit volume increases and the probability for  finding at least one of them around $100\, \text{pc}$ is practically 1.}. In addition, there is around $20\%$ of chance that at least one of these dressed PBH is located within the detectable distance (where Eq.~(\ref{fluxps}) equals the HESS sensitivity)  via $\gamma$-ray (line and continuum) emission.  Interesting enough, these values are comparable with the $30\%$ of chances of finding one WIMPs ultracompact minihalo grown during the $e^+-e^-$ annihilation epoch from a density perturbation of $1/3 M_{\odot}$~\cite{Scott:2009tu}. 

We mention that assumptions taken in Eqs.~(\ref{fPBHdbar}) and (\ref{probone})
are not robust against the presence of local DM substructure in the solar vicinity.
N-body simulations aimed to understand the granularity of DM halos have shown significant variations in density over a sphere of hundreds parsecs~\cite{Zemp:2008gw}. Thus, our 
constraints derived from the nearest dressed PBH as well as probabilities for 
finding these compact objects at certain detectable distance should be taken with caution.  

\section{Velocity of Particles}
 
So far, we have assumed the wino DM particles are non-relativistic. In particular, the cross section $\langle\sigma v\rangle_{\rm line}$ used here is derived assuming the typical velocities do not exceed  $\mathcal{O}(10^{-3})\,\text{c}$ \cite{Ovanesyan:2014fwa}. We now discuss the validity of this assumption. 

We may estimate the velocity dependence of DM particles in minihalos around PBHs by considering the virial velocity
as 
\begin{align}
v_{\chi^0}(r) & \simeq \left[ \frac{G_N (M^{\text{PBH}}_{\text{halo}}(r)+M_{\text{PBH}})}{r} \right]^{1/2}\,\nonumber\\
&\simeq \left[ \frac{1}{2\tilde{r}} + \frac{2\pi r_g^3 \rho_{\text{max}}\tilde{r}^2}{3 M_{\text{PBH}}} \right]^{1/2}\,,
\end{align}
where he have used Eq.~(\ref{eq:rhoapp}) and concentrated on distances $r < r^*$. Suppose that at a radial distance $\tilde{r} < \tilde{r}_{\text{rel}}$ the
wino velocity is $v_{\chi^0}(\tilde{r}_{\text{rel}}) > 10^{-3}c$. Since the photon line annihilation rate scales with the radius as $\Gamma_{\text{line}}^{\text{PBH}} \sim \tilde{r}^3$, we may estimate the fraction of the  line annihilation rate in the inner parts of minihalos with respect to the total 
rate. For example, for the case of PBHs with masses in the range $10^{-12}\,M_{\odot}\lesssim M_{\text{PBH}} \lesssim M_{\odot}$, we have
\begin{equation}
\mathcal{O}(10)^{-31}\lesssim \frac{\Gamma_{\text{line}}^{\text{PBH}}(\tilde{r}_{\text{rel}})}{\Gamma_{\text{line}}^{\text{PBH}}(\tilde{r}^*)}\lesssim \mathcal{O}(10^{-7})\,,     
\end{equation}
so that our analysis hold in a good approximation.

\section{Summary}

In this study we have studied the viability of the well-motivated mixed DM scenario composed of
a dominant thermal WIMP, with a focus on $SU(2)_L$ triplet fermion ``winos", and a small fraction in PBHs.  After the wino kinetic decoupling, the wino particles are gravitationally captured by PBHs so that we expect today the presence of PBHs with dark minihalos in the Milky Way. Even though the wino annihilation is enhanced in such compact astrophysical objects, by using the H.E.S.S. data for $\gamma$-ray lines from the Galactic Center, we have shown that the scenario is viable for sufficiently small fraction of PBHs in a Milky Way with a DM  cored halo profile of some kpc. Without considering the granularity of the DM halo in the solar vicinity, for the case of dressed PBHs having a central PBH with $M_{\text{PBH}} \sim M_{\odot}$, we find a sizeable chance for observation on the Earth by present or upcoming experiments. This would be an observation of a single (primordial) black hole in the sky acting as a kind of light source; this would be a spectacular way to discover new physics.

Since formation of minihalos with density spikes around PBHs and annihilation
of DM particles in astrophysical compact objects are both quite general processes, 
our work can be readily generalized to other setups, including other DM candidates and/or ultracompact minihalos.

\section{Acknowledgments}
E. D. S. thanks Rong-Gen Cai, Xing-Yu Yang and Yu-Feng Zhou for useful discussions about
dark matter profiles around PBHs. E. D. S. thanks Luca Visinelli for useful discussion about indirect DM searches. T. T. Y. thanks Satoshi Shirai for useful discussion on the formation of dressed PBHs.
This work was supported by the Academy of Finland grant 318319.
 T. T. Y. is supported in part by the China Grant for Talent Scientiﬁc
Start-Up Project and the JSPS Grant-in-Aid for Scientiﬁc Research Grants No.
16H02176, No. 17H02878, and No. 19H05810 and by World Premier
International Research Center Initiative (WPI Initiative), MEXT, Japan.
M. P. H. is supported in part 
by National Science Foundation Grant No. PHY-2013953.\\
$^*$\href{mailto:mark.hertzberg@tufts.edu}{mark.hertzberg@tufts.edu}\\
$^{**}$\href{sami.t.nurmi@jyu.fi}{sami.t.nurmi@jyu.fi}\\
$^\dagger$\href{mailto:edschiap@uc.cl}{edschiap@uc.cl}\\
$\ddagger$\href{mailto:tsutomu.tyanagida@ipmu.jp}{tsutomu.tyanagida@ipmu.jp}

\bibliography{ShiningPBHs} % Tell bibtex which .bib file to use (this one is some example file in TexLive's file tree)

\end{document}